\documentclass[12pt]{article}

\catcode`\@=11
\@addtoreset{equation}{section}

\global\arraycolsep=2pt

\usepackage{mathrsfs,amsbsy,amssymb,latexsym,amsfonts,amsmath,cite}
\usepackage{graphicx,color}
\usepackage{mathtools}

\allowdisplaybreaks

\newcommand{\al}{\alpha}
\newcommand{\be}{\beta}
\newcommand{\ga}{\gamma}
\newcommand{\de}{\delta}
\newcommand{\ep}{\epsilon}

\newcommand{\ze}{\zeta}

\newcommand{\Tr}{{\rm Tr}}
\newcommand{\str}{{\rm STr}}

\newcommand{\alg}[1]{\mathfrak{#1}}
\newcommand{\el}{\nonumber}
\newcommand{\nln}{\nonumber\\}

\begin{document}

\begin{flushright}
\parbox{4cm}
{KUNS-2543}
\end{flushright}

\vspace*{1.5cm}

\begin{center}
{\Large\bf Schr\"odinger geometries arising from \\ Yang-Baxter deformations} 
\vspace*{1.5cm}\\
{\large 
Takuya Matsumoto$^{\dagger}$\footnote{E-mail:~takuya.matsumoto@math.nagoya-u.ac.jp} 
and Kentaroh Yoshida$^{\ast}$\footnote{E-mail:~kyoshida@gauge.scphys.kyoto-u.ac.jp}} 
\end{center}
\vspace*{0.25cm}
\begin{center}
$^{\dagger}${\it 
Graduate School of Mathematics and Institute for Advanced Research, \\  
Nagoya University, Nagoya 464-8602, Japan. }
\vspace*{0.25cm}\\ 
$^{\ast}${\it Department of Physics, Kyoto University, \\ 
Kyoto 606-8502, Japan.} 
\end{center}
\vspace{1cm}

\begin{abstract}
We present further examples of the correspondence between solutions of type IIB supergravity 
and classical $r$-matrices satisfying the classical Yang-Baxter equation (CYBE). 
In the previous works, classical $r$-matrices have been composed of generators of 
only one of either $\mathfrak{so}(2,4)$ or $\mathfrak{so}(6)$\,. 
In this paper, we consider some examples of $r$-matrices with both of them. 
The $r$-matrices of this kind contain (generalized) Schr\"odinger spacetimes and gravity duals of 
dipole theories. It is known that 
the generalized Schr\"odinger spacetimes can also be obtained via a certain class of TsT transformations 
called null Melvin twists. The metric and NS-NS two-form are reproduced by following 
the Yang-Baxter sigma-model description. 
\end{abstract}

\setcounter{footnote}{0}
\setcounter{page}{0}
\thispagestyle{empty}

\newpage
\tableofcontents

\section{Introduction} 

The gauge/gravity correspondence provides a fascinating 
arena for studying new aspects of string theory. 
A well-studied example is the duality between type IIB superstring 
theory on AdS$_5\times$S$^5$ and the $\mathcal{N}$=4 super Yang-Mills (SYM) 
theory in four dimensions \cite{M}. This duality enjoys a powerful property, 
integrability (For a big review, see \cite{review}).  
It enables one to compute some physical quantities exactly 
even at finite coupling without supersymmetries. Thus the integrable structure behind the duality 
will be more and more important in the future study. 

\medskip 

The integrable structure may provide another insight as well.  
It is integrable deformations of the AdS/CFT correspondence. 
It is well recognized that type IIB string theory on AdS$_5\times$S$^5$ \cite{MT}, 
which is often called the AdS$_5\times$S$^5$ superstring, 
is classically integrable \cite{BPR}. 
A systematic way to study the deformations is to employ the Yang-Baxter sigma-model formulation, 
which was originally proposed by Klimcik \cite{Klimcik} for principal chiral models\footnote{
For the $SU(2)$ case, a $q$-deformed algebra and its affine extension have been revealed 
in \cite{KY,KYhybrid,KMY-QAA}.}
and generalized by Delduc-Magro-Vicedo \cite{DMV} to coset sigma models. 
In this formalism, a deformed target space is determined by specifying 
the associated classical $r$-matrix satisfying the modified classical Yang-Baxter equation 
(mCYBE). This formalism has been extended to the case with the Wess-Zumino-Witten (WZW)-like term 
\cite{DMV-WZW}\footnote{For the related progress on integrability with the WZW-term, see also \cite{KOY,S-lambda}.}. 
One may consider a generalization of it to the standard classical 
Yang-Baxter equation (CYBE) \cite{MY-YBE}. A three-dimensional Schr\"odinger spacetime is 
contained here as a simple case with the rank 1 and for this case the classical integrable 
structure has been recognized as Jordanian twists \cite{KY-Sch}\footnote{
It may be significant to unveil the relation between the classical $r$-matrix 
and the general invariant two-form in the coset construction \cite{SYY}. 
It is also interesting to figure out the relation to the construction \cite{ORU}.}.  

\medskip 

The formalism of the Yang-Baxter deformations is applicable 
to the AdS$_5\times$S$^5$ superstring \cite{DMV2}. 
Then it has been generalized to the CYBE case \cite{KMY-Jordanian-typeIIB}. 
In comparison to the mCYBE, there are two advantages. 
The first is that partial deformations of AdS$_5\times$S$^5$ are possible. 
For example, one may consider a deformation of either AdS$_5$ or S$^5$\,, 
while it seems likely, so far as we see the current achievement, 
that both of them are inevitably deformed in the case of mCYBE \cite{DMV2}. 
The second is that there are a variety of solutions of the CYBE. In fact, 
we have already found various examples of classical $r$-matrices 
which correspond to solutions of type IIB supergravity. 
For example, the Lunin-Maldacena-Frolov background \cite{LM,Frolov} 
and the gravity duals of non-commutative gauge theories \cite{HI,MR} 
have been reproduced in \cite{LM-MY} and \cite{MR-MY}, respectively, 
by following the Yang-Baxter sigma-model description.  
Thus, the correspondence of this kind may be called the gravity/CYBE correspondence 
\cite{LM-MY} (For a short summary see, \cite{MY-summary}). 

\medskip 

A remarkable point of the gravity/CYBE correspondence 
is the connection to TsT transformations \cite{MY-duality,SUGRA-KMY}. That is, 
a certain class of classical $r$-matrices satisfying the CYBE 
is related to a solution generation techniques in type IIB supergravity. 
More strikingly, it is shown in \cite{CMY} that this connection holds even for Yang-Baxter 
deformations of a non-integrable background, AdS$_5\times T^{1,1}$ \cite{BZ}. 
Motivated by these developments, 
our purpose here is to pursuit the relation to TsT transformations and 
discover further examples of the gravity/CYBE correspondence. 

\medskip 

Indeed, we will present further examples of the gravity/CYBE correspondence. 
In the previous works, classical $r$-matrices have been composed of generators of 
only one of either $\mathfrak{so}(2,4)$ or $\mathfrak{so}(6)$\,. 
In the preset paper, we consider some examples of $r$-matrices with both of them, 
which are schematically of the form; 
\begin{align}
r= \sum_i (a_i\otimes b_i-b_i\otimes a_i)
\qquad\text{with}\qquad
a_i\in \alg{so}(2,4)\,,\quad b_i\in \alg{so}(6)\,. 
\label{intro-r}
\end{align}
The classical $r$-matrices of this kind contain (generalized) 
Schr\"odinger spacetimes and 
gravity duals of dipole theories. It is known that the generalized Schr\"odinger spacetimes 
can also be obtained via a certain class of TsT transformations called null Melvin twists 
as discussed in \cite{MMT} and \cite{BK,BKP}. The metric and NS-NS two-form are reproduced by following 
the Yang-Baxter sigma-model description. 

\medskip 

This paper is organized as follows. 
Section \ref{sec:2} gives a short review of Yang-Baxter deformations of the
AdS$_5\times$S$^5$ superstring based on the CYBE. 
Section \ref{sec:3} presents classical $r$-matrices associated with (generalized) 
Schr\"odinger spacetimes. 
Section \ref{sec:4} considers dipole deformations of the AdS$_5\times$S$^5$ background 
with the corresponding $r$-matrices. 
Section \ref{sec:5} is devoted to conclusion and discussion. 
In Appendix \ref{app:notation}, we summarize our convention and notation 
of the $\mathfrak{so}(2,4)$ and $\mathfrak{so}(6)$ generators. 
In Appendix \ref{app:Tdual}, the T-duality rules are listed. 
Appendix \ref{app:C} gives a derivation of three-parameter dipole deformations 
of AdS$_5\times$S$^5$\,.

\section{Yang-Baxter deformations of AdS$_5\times$S$^5$}
\label{sec:2}

Let us first recall the formulation of Yang-Baxter sigma models. 

\medskip

The Yang-Baxter sigma-model description was originally developed 
for purely bosonic non-linear sigma models \cite{Klimcik,DMV}. 
It is now generalized to supersymmetric cases, 
and integrable deformations of the AdS$_5\times$S$^5$ superstring can be 
described based on the mCYBE \cite{DMV2} and the CYBE \cite{KMY-Jordanian-typeIIB}. 

\medskip 

Here we consider the latter case \cite{KMY-Jordanian-typeIIB}, 
where the deformed classical action is given by 
\begin{align}
S=-\frac{1}{4}(\ga^{\al\be}-\ep^{\al\be}) 
\int^\infty_{-\infty}\!\! d\tau \int^{2\pi}_{0}\!\! d\sigma\,   
\str\Bigl(A_\al d\circ \frac{1}{1-\eta R_g\circ d}A_\be \Bigr)\,, 
\label{action}
\end{align}
where the left-invariant one-form is defined as  
\begin{align}
A_\al\equiv g^{-1} \partial_\al g\,, \qquad  g\in SU(2,2|4)\,.  
\end{align}
When $\eta = 0$\,, the classical action \eqref{action} is reduced to the undeformed one\cite{MT}. 
Here the string world-sheet metric is taken to be flat i.e., 
$\ga^{\al\be} = {\rm diag}(-1,1)$\,. The anti-symmetric tensor 
$\ep^{\al\be}$ is normalized as $\ep^{\tau\sigma}=1$\,. 

\medskip 
 
The operator $R_g$ is defined as 
\begin{align}
R_g(X)\equiv g^{-1}R(gXg^{-1})g\,, 
\end{align}
where a linear operator $R$ is a solution of CYBE rather than mCYBE.  
The R-operator is related to a classical $r$-matrix in the tensorial notation through 
\begin{align}
&R(X)=\str_2[r(1\otimes X)]=\sum_i \bigl(a_i\str(b_iX)-b_i\str(a_iX)\bigr) 
\label{linearR} \\
&\text{with}\quad r=\sum_i a_i\wedge b_i\equiv \sum_i (a_i\otimes b_i-b_i\otimes a_i)\,. \el 
\end{align}
The generators $a_i, b_i$ are some elements of $\alg{su}(2,2|4)$\,. 
The supertrace $\str$ of $(4|4)\times(4|4)$ supermatrix is defined as 
\begin{align}
\str\left(\begin{array}{c|c}
A & B \\ \hline C & D 
\end{array}\right)
=\Tr(A)-\Tr(D)\,, 
\end{align}
where each of the blocks $A,B,C,D$ is a $4\times4$ matrix of complex numbers, 
which plays a crucial role in our argument.

\medskip 

Note that $\alg{su}(2,2|4)$ enjoys the $\mathbb{Z}_4$-grading property 
and one can introduce the projectors $P_k$ ($k=0,1,2,3$) from $\alg{su}(2,2|4)$
to its $\mathbb{Z}_4$-graded components $\alg{su}(2,2|4)^{(k)}$\,. 
In particular, $\alg{su}(2,2|4)^{(0)}$ is a gauge symmetry, $\alg{so}(1,4)\oplus\alg{so}(5)$\,. 
Then the operator $d$ is defined as a linear combination of $P_k$\,, 
\begin{align}
d\equiv P_1+2P_2-P_3\,.   
\end{align} 
The numerical coefficients are fixed by requiring the kappa-symmetry \cite{KMY-Jordanian-typeIIB}.

\medskip 

To evaluate the action, it is convenient to rewrite the metric part and NS-NS two-form coupled part 
of the Lagrangian \eqref{action} into the following form,  
\begin{align}
L_G&=\frac{1}{2}\str\left[A_\tau P_2(J_\tau)-A_\sigma P_2(J_\sigma)\right]\,, \el \\
L_B&=\frac{1}{2}\str\left[A_\tau P_2(J_\sigma)-A_\sigma P_2(J_\tau)\right]\,, 
\label{LGLB} 
\end{align} 
where $J_\al$ is a projected current defined as 
\begin{align}
J_\al \equiv \frac{1}{1-2\eta R_g\circ P_2}A_\al \,. 
\label{bos-J}
\end{align}

\medskip 

One can read off the deformed metric  and 
NS-NS two-form from $L_G$ and $L_B$\,, respectively. 
It is quite messy but straightforward. 
It would be an easy exercise to derive the metric and NS-NS two-form 
by following the previous examples \cite{SUGRA-KMY,LM-MY,MR-MY,MY-summary,MY-duality}.

\section{Integrability of Schr\"odinger spacetimes}
\label{sec:3}

In this section, we will deduce Shcr\"odinger spacetimes from classical 
$r$-matrices depending on both $\alg{so}(2,4)$ and $\alg{so}(6)$\,. 
Consequently, the classical integrability of the spacetimes  
automatically follows from the Yang-Baxter sigma model formulation. 
After presenting the well-known Shcr\"odinger spacetime in
subsection \ref{subsec:3.1}\,, we consider three-parameter generalizations in 
subsection \ref{subsec:3.2}\,.  
More complicated examples are presented in subsection \ref{subsec:3.3}\,.  

\subsection{Schr\"odinger geometries as Yang-Baxter deformations} 
\label{subsec:3.1} 

Inspired by the spirit of the gravity/CYBE correspondence, 
we have found that the following classical $r$-matrix 
\begin{align}
r=-\frac{i\be}{4\eta}\,  p_- \wedge (h_4 +h_5+h_6)
\label{r-ma}
\end{align}
corresponds to the Schr\"odinger background. 
Here $p_-$ is the light-cone generator in $\alg{so}(2,4)$ and 
$h_4,h_5$ and $h_6$ are the Cartan generators in $\alg{so}(6)$\,. 
Hence the $r$-matrix in (\ref{r-ma}) is indeed of the form \eqref{intro-r}. 
The parameter $\be$ measures the deformation\footnote{
The parameter $\eta$ is not an essential deformation parameter 
because it is canceled in \eqref{bos-J}\,.}.
For the details of our convention and notation of the generators, 
see Appendix \ref{app:notation}. 

\medskip 

To find the metric and NS-NS two-form from \eqref{LGLB}\,, 
we need to evaluate the projected deformed current $P_2(J_\al)$\,. 
By solving the equation
\begin{align}
(1-2\eta P_2 \circ R_g )P_2(J_\al) =P_2(A_\al) \,, 
\end{align}
which is obtained from the definition \eqref{bos-J}\,, 
the deformed current is evaluated as  
\begin{align}
P_2(J_\al)&=c^1\ga^a_1+c^2\ga^a_2+c^3\ga^a_3+c^0\ga^a_0+c^5\ga^a_5 \nln
& \quad  +d^1\ga^s_1+d^2\ga^s_2+d^3\ga^s_3+d^4\ga^s_4+d^5\ga^s_5\,, 
\end{align}
with the coefficients\footnote{For the conventions of the gamma-matrices, 
see Appendix \ref{app:notation}.}
\begin{align}
c^1&=\frac{\partial_\al x^1}{2z}\,, \qquad 
c^2=\frac{\partial_\al x^2}{2z}\,, \qquad 
c^5=\frac{\partial_\al z}{2z}\,, \nln
c^3&=-\frac{\be^2\partial_\al x^+}{2\sqrt{2}z^3}
+\frac{1}{2\sqrt{2}z}\Bigl(\partial_\al x^+ -\partial_\al x^-
+\be\Bigl(\partial_\al\chi
+\frac{1}{2}\sin^2\mu(\partial_\al\psi+\cos\theta\partial_\al\phi)\Bigr)\Bigr)\,,\nln
c^0&=+\frac{\be^2\partial_\al x^+}{2\sqrt{2}z^3}
+\frac{1}{2\sqrt{2}z}\Bigl(\partial_\al x^+ +\partial_\al x^-
-\be\Bigl(\partial_\al\chi
+\frac{1}{2}\sin^2\mu(\partial_\al\psi+\cos\theta\partial_\al\phi)\Bigr)\Bigr)\,,\nln
d^1&=-\frac{i}{2}\partial_\al \mu\,, \qquad 
d^3=-\frac{i}{4}\sin\mu\,\partial_\al\theta\,, \qquad 
d^5=\frac{i}{2z^2}\cos\mu (z^2\partial_\al\chi-\be\partial_\al x^+)\,, 
\nln
d^2&=\frac{i}{4z^2}\cos\frac{\theta}{2} \sin\mu \bigl(
2\be\partial_\al x^+
-z^2 (\partial_\al \phi+2\partial_\al \chi+\partial_\al \psi )\bigr)\,,\nln
d^4&=\frac{i}{4z^2}\sin\frac{\theta}{2} \sin\mu \bigl(
2\be\partial_\al x^+
+z^2 (\partial_\al \phi-2\partial_\al \chi-\partial_\al \psi )\bigr)\,. 
\end{align}

\medskip 

Plugging the above expression of $P_2(J_\al)$ with \eqref{LGLB}\,, 
the resulting metric and NS-NS two-form turn out to be  
\begin{align}
ds^2&=\frac{-2dx^+dx^-+(dx^1)^2+(dx^2)^2+dz^2}{z^2}
- \be^2 \frac{(dx^+)^2}{z^4} +ds^2_{\rm S^5}\,, \nln 
B_2&=\frac{\be}{z^2}dx^+\wedge (d\chi+\omega)\,.  
\label{sch}
\end{align}
Here the line element $ds^2_{\rm S^5}$ is measured by the metric of S$^5$ 
with the coordinates $(\chi,\mu,\psi,\theta,\phi)$\,,  
\begin{align} 
ds^2_{\rm S^5}&=(d\chi+\omega)^2 +ds^2_{\rm \mathbb{C}P^2}\,, \nln 
ds^2_{\rm \mathbb{C}P^2}&= d\mu^2+\sin^2\mu\,
\bigl(\Sigma_1^2+\Sigma_2^2+\cos^2\mu\,\Sigma_3^2\bigr)\,.  
\label{S1overCP2}
\end{align}
Namely, the round S$^5$ is expressed as an S$^1$-fibration over $\mathbb{C}$P$^2$\,,  
where $\chi$ is the fiber coordinate and $\omega$ is 
a one-form potential of the K\"ahler form on $\mathbb{C}$P$^2$\,. 
The symbols $\Sigma_i ~(i=1,2,3)$ and $\omega$ are defined as  
\begin{align}
\Sigma_1&= \tfrac{1}{2}(\cos\psi\, d\theta +\sin\psi\sin\theta\, d\phi)\,, \nln 
\Sigma_2&= \tfrac{1}{2}(\sin\psi\, d\theta -\cos\psi\sin\theta\, d\phi)\,, \nln 
\Sigma_3&= \tfrac{1}{2}(d\psi +\cos\theta\, d\phi)\,, 
\qquad 
\omega=\sin^2\mu\, \Sigma_3\,. 
\end{align} 

\medskip 

The metric of the deformed AdS$_5$ part is nothing but the metric that was 
originally proposed in \cite{Son,BM}. This metric preserves a non-relativistic 
conformal symmetry called the Schr\"odinger symmetry \cite{Sch}. 
This metric can be reproduced via a coset construction \cite{SYY}. 

\medskip 

As a result, we have proven the classical integrability of the string sigma model 
whose target space is given by the Schr\"odinger background \eqref{sch} 
in the sense that the Lax pair has been constructed\footnote{
Note here that non-integrability of various non-relativistic backgrounds was argued in \cite{NR-Sfe}, 
but the Schr\"odinger spactime with the dynamical critical exponent $z=2$ has not been covered. 
Hence, our result is not in contradiction with \cite{NR-Sfe}.}.

\subsection{Three-parameter generalizations} 
\label{subsec:3.2}

In the previous subsection, we have considered a one-parameter deformation. 
Then, one may consider multi-parameter deformations. 
An example of three-parameter deformation is described by the 
following classical $r$-matrix,  
\begin{align}
r(\vec\be)=-\frac{i}{4\eta}\, p_- \wedge (\be_1\,h_4 +\be_2\,h_5+\be_3\,h_6)\,,
\end{align}
where $\vec\be\equiv(\be_1,\be_2,\be_3)$ are the real deformation parameters. 

\medskip 

Since the calculations to find the metric and NS-NS two-form
are completely parallel to the previous section, 
we do not repeat them here. 
With the S$^5$ metric \eqref{app:S1overCP2}, 
the final expressions are written as 
\begin{align}
ds^2&=\frac{-2dx^+dx^-+(dx^1)^2+(dx^2)^2+dz^2}{z^2}
-\frac{f(\vec\be)(dx^+)^2}{4z^4} +ds^2_{\rm S^5}\,, \nln 
B_2&=\frac{1}{4z^2}dx^+\wedge K(\vec\be)\,,  
\end{align}
where $f(\vec\be)$ and $K(\vec\be)$ are a scalar function and a one-form on 
$S^5$\,, respectively, depending on $\vec\be$\,.
They are explicitly defined as 
\begin{align}
f(\vec\be)
&=\be_1^2+\be_2^2+2\be_3^2
+2 (\be_1^2-\be_2^2)\cos\theta\sin^2\mu
-\left(\be_1^2+\be_2^2-2\be_3^2\right) \cos2\mu \,, 
\nln 
K(\vec\be)
&=\left(2 (\be_1-\be_2) \cos\theta\sin^2\mu-(\be_1+\be_2-2\be_3) 
\cos2\mu+\be_1+\be_2+2\be_3 \right)d\chi \nln
&\quad +(\be_1+\be_2+(\be_1-\be_2) \cos\theta)\sin^2\mu d\psi \nln 
&\quad +(\be_1-\be_2+(\be_1+\be_2) \cos\theta)\sin^2\mu d\phi\,. 
\end{align}

\medskip

The above expressions are quite messy and it is convenient to rewrite them 
in terms of another coordinate system of S$^5$\,. 
In fact, by following the argument in Appendix \ref{app:notation},  
one can rewrite them into the following simpler form:   
\begin{align}
ds^2&=\frac{-2dx^+dx^-+(dx^1)^2+(dx^2)^2+dz^2}{z^2}
- \Bigl(\sum_{i=1}^3\be_i^2\, \mu_i^2\,\Bigr)
\frac{(dx^+)^2}{z^4} +ds^2_{\rm S^5}\,,
\nln 
B_2&=\frac{1}{z^2}\,dx^+\wedge \Bigl(\sum_{i=1}^3\be_i\, \mu_i^2\,d\psi_i\Bigr) \,,  
\qquad \mu_1^2+\mu_2^2+\mu_3^2=1\,. \label{BK}
\end{align}
Here $\mu_1$\,, $\mu_2$ and $\mu_3$ are the S$^5$ coordinates defined as 
\begin{align}
\mu_1=\cos\ze\, \sin r \,, \qquad 
\mu_2=\sin\ze\, \sin r \,, \qquad 
\mu_3=\cos r\,. 
\label{mu123}
\end{align}
This background (\ref{BK}) nicely agrees with the one obtained by Bobev and Kundu \cite{BK} 
via null Melvin twists of AdS$_5\times$S$^5$\,. 
As a result, this background (\ref{BK}) also gives rise to an integrable 
background of type IIB string theory. A remarkable point is that 
the deformation term of the AdS$_5$ part depends on the S$^5$ coordinates. 
This dependence does not break the Schr\"odinger symmetry. 
It can be generalized beyond the TsT transformations. 
For example, the dependence of this type has been studied in \cite{HY}. 
It would be interesting to study whether the classical integrability is 
preserved for the deformation argued in \cite{HY}. 
The stability of the solution might be related to the classification 
of possible classical $r$-matrices.

\medskip 

Note that when the deformation parameters take the special values,  
\begin{align}
\be_1=\be_2=\be_3=\be\,,  
\end{align}
the background (\ref{BK}) reduces to the Schr\"odinger geometry \eqref{sch} 
by taking account of the relation, 
\begin{align}
d\chi+\omega=\mu_1^2d\psi_1+\mu_2^2d\psi_2+\mu_3^2d\psi_3\,. 
\end{align}

Finally it is worth noting that the Schr\"odinger spacetimes discussed so far 
are non-supersymmetric for generic values. 
This is basically because T-dualities are taken for R-symmetry directions 
and then all of the supersymmetries are broken. 
For supersymmetric Schr\"odinger backgrounds, see the following subsection.

\subsection{Other examples}
\label{subsec:3.3} 

It would be worth giving more examples. 
We will give five examples of classical $r$-matrices satisfying the CYBE. 
The $r$-matrices always contain $p_-$ from $\mathfrak{so}(2,4)$, 
and the other generators are picked up from $\mathfrak{so}(6)$\,,
which are not the Cartan generators. 

\medskip

To present the examples, 
the following polar coordinates of S$^5$ are more convenient:  
\begin{align}
n_1 &=\mu_1 \cos\psi_1\,, \qquad n_2  = \mu_1 \sin\psi_1\,, \nonumber \\
n_3 &=\mu_2 \cos\psi_2\,, \qquad n_4 = \mu_2 \sin\psi_2\,, \nonumber \\ 
n_5 &=\mu_3 \cos\psi_3\,, \qquad n_6 = \mu_3 \sin\psi_3\,, \\
\text{where}\qquad 
\mu_1& = \cos\zeta\sin r\,, \qquad 
\mu_2 = \sin\zeta\sin r\,, \qquad 
\mu_3 = \cos r\,. \el 
\end{align}
The above coordinates satisfy the following condition: 
\begin{align}
(n_1^2+n_2^2)+(n_3^2+n_4^2)+(n_5^2+n_6^2)
=\mu_1^2+\mu_2^2+\mu_3^2=1\,.
\end{align}

\subsubsection*{1) \quad classical $r$-matrix with $\ga^s_{1}$ }

The first examples is a classical $r$-matrix composed of 
$p_-$ and $\ga^s_1$ like 
\begin{align}
r=-\frac{i}{4}\, p_- \wedge \ga^s_1\,. 
\end{align}
This $r$-matrix gives rise to the following metric and NS-NS two-form: 
\begin{align}
ds^2&=ds^2_{\rm AdS_5}+ds^2_{\rm S^5}
+\eta^2(\mu_1^2\cos^2\psi_1+\mu_3^2\cos^2\psi_3)\frac{(dx^+)^2}{z^4}
\nln 
&=ds^2_{\rm AdS_5}+ds^2_{\rm S^5} +\eta^2(n_1^2+n_5^2 )\frac{(dx^+)^2}{z^4}\,,
\nln 
B_2&=
\frac{\eta }{z^2}\, dx^+\wedge 
\bigl((\mu_1\cos\psi_1) \, d(\mu_3\cos\psi_3)
- (\mu_3\cos\psi_3) \, d(\mu_1\cos\psi_1) \bigr)\,. 
\nln 
&=\frac{\eta }{z^2}\, dx^+\wedge (n_1dn_5 - n_5dn_1) \,. 
\end{align}

\subsubsection*{2) \quad classical $r$-matrix with $n_{13}$ }

The second example is a classical $r$-matrix, 
\begin{align}
r=\frac{1}{2}\, p_- \wedge n_{13}\,. 
\end{align}
The resulting metric and NS-NS two-form are given by 
\begin{align}
ds^2&=ds^2_{\rm AdS_5}+ds^2_{\rm S^5}
-\eta^2(\mu_1^2\cos^2\psi_1+\mu_2^2\cos^2\psi_2)\frac{(dx^+)^2}{z^4} 
\nln
&=ds^2_{\rm AdS_5}+ds^2_{\rm S^5}
-\eta^2(n_1^2+n_3^2)\frac{(dx^+)^2}{z^4}\,,
\nln  
B_2&= \frac{\eta }{z^2}\, dx^+\wedge 
\bigl((\mu_1\cos\psi_1) \,d(\mu_2\cos\psi_2)
- (\mu_2\cos\psi_2) \,d(\mu_1\cos\psi_1) \bigr)
\nln 
&= \frac{\eta }{z^2}\, dx^+\wedge (n_1dn_3-n_3dn_1) \,. 
\end{align}

\subsubsection*{3) \quad classical $r$-matrix with $n_{15}$ }

The third example is a classical $r$-matrix, 
\begin{align}
r=-\frac{1}{2}\, p_- \wedge n_{15}\,. 
\end{align}
The resulting metric and NS-NS two-form are given by 
\begin{align}
ds^2&=ds^2_{\rm AdS_5}+ds^2_{\rm S^5}
-\eta^2(\mu_1^2\cos^2\psi_1+\mu_3^2\sin^2\psi_3)\frac{(dx^+)^2}{z^4} 
\nln
&=ds^2_{\rm AdS_5}+ds^2_{\rm S^5} -\eta^2(n_1^2+n_6^2)\frac{(dx^+)^2}{z^4}\,,
\nln  
B_2&= \frac{\eta }{z^2}\, dx^+\wedge 
\bigl((\mu_1\cos\psi_1) \,d(\mu_3\sin\psi_3)
- (\mu_3\sin\psi_3) \,d(\mu_1\cos\psi_1) \bigr) \nln 
&= \frac{\eta }{z^2}\, dx^+\wedge (n_1dn_6-n_6dn_1 )\,.  
\end{align}

\subsubsection*{4) \quad classical $r$-matrix with $n_{12}$\,, $n_{23}$ and $n_{34}$ }

The fourth example is a classical $r$-matrix,  
\begin{align}
r=\frac{1}{4}\, p_- \wedge (\al_1n_{12}+\al_2 n_{23}+\al_3 n_{34})\,. 
\end{align}
The resulting metric and NS-NS two-form are given by 
\begin{align}
ds^2&=ds^2_{\rm AdS_5}+ds^2_{\rm S^5} \nln 
& \quad 
-\eta^2\bigl(\al_1^2(n_1^2+n_2^2)+\al_2^2(n_2^2+n_3^2)+\al_3^2(n_3^2+n_4^2)
-2 \al_2 (\al_1n_1n_3 +\al_3 n_2n_4 )\bigr) \frac{(dx^+)^2}{z^4}\,,
\nln 
B_2&= \frac{\eta }{z^2}\, dx^+\wedge 
\bigl(\al_1(n_1dn_2-n_2dn_1)+\al_2(n_2dn_3-n_3dn_2)+\al_3(n_3dn_4-n_4dn_3) \bigr)\,. 
\end{align}

\subsubsection*{5) \quad classical $r$-matrix with $n_{12}$\,, $n_{34}$ and $h_6$ }

The fifth example is a classical $r$-matrix,  
\begin{align}
r=\frac{1}{4}\, p_- \wedge \Bigl(\al_1n_{12}+\al_2 n_{34}-\frac{i\al_3}{2} h_6\Bigr)\,. 
\label{BKP-r}
\end{align}
Here it is noted that three $\alg{so}(6)$ generators 
$n_{12}\,, n_{34}$ and $h_6$ commute each other. 
The resulting metric and NS-NS two-form are given by 
\begin{align}
ds^2&=ds^2_{\rm AdS_5}+ds^2_{\rm S^5} 
-\eta^2\bigl(\al_1^2(n_1^2+n_2^2)+\al_2^2(n_3^2+n_4^2)+\al_3^2(n_5^2+n_6^2)
\bigr) \frac{(dx^+)^2}{z^4}\,,
\nln 
B_2&= \frac{\eta }{z^2}\, dx^+\wedge 
\bigl(\al_1(n_1dn_2-n_2dn_1) \nln
& \hspace*{2.5cm}  +\al_2(n_3dn_4-n_4dn_3)+\al_3(n_5dn_6-n_6dn_5) \bigr)\,. 
\label{BKP}
\end{align}
It should be remarked that the background with (\ref{BKP}) 
agrees with the one found in \cite{BKP}. 
Thus the existence of the associated classical $r$-matrix (\ref{BKP-r}) indicates that 
the string theory defined on the background \cite{BKP} 
is classically integrable in the sense that the Lax pair is constructed. 

\medskip 

According to the Killing spinor analysis in \cite{BKP}, for values satisfying the condition 
\begin{equation}
\alpha_1 \pm \alpha_2 \pm \alpha_3 =0\,, 
\label{susy}
\end{equation}
two real supersymmetries are preserved. 
When, in addition to the condition (\ref{susy}), at least one of the $\alpha_i~(i=1,2,3)$ vanishes, 
four real supersymmetries are preserved. 
Thus, depending on classical $r$-matrices, the remaining supersymmetries should be different. 
Hence it would be interesting to study the relation between classical $r$-matrices 
and the classification of super Schr\"odinger algebras \cite{SY-Sch}. It would also be 
nice to study the relation to warped AdS$_3$ geometries, for example, along the line of \cite{JOY}.

\section{Integrability of gravity duals for dipole theories}
\label{sec:4}

In this section, we shall derive gravity duals of dipole theories \cite{dipole1,dipole2,dipole3,dipole4}  
from the viewpoint of the Yang-Baxter deformations. 
We find out classical $r$-matrices associated with the dipole backgrounds 
and derive the deformed metric and NS-NS two-form both for a one-parameter case 
and a three-parameter case, 
in subsection \ref{subsec:4.1} and \ref{subsec:4.2}\,, respectively.

\subsection{A one-parameter case}
\label{subsec:4.1} 

Let us first consider a one-parameter dipole background. 
It can be obtained by a TsT-transformation $(x^3, \psi_1)_{\al_1}$\,, 
where $\al_1$ is a shift parameter $\al_1$ \cite{Imeroni}. 
From this information on the TsT-transformation, one can easily guess 
the corresponding classical $r$-matrix based on the knowledge obtained \cite{MY-duality}. 

\medskip 

A natural candidate of the associated $r$-matrix is given by  
\begin{align}
r=\frac{i\al_1 }{4\eta}\, p_3\wedge h_4\,,  \label{1-para-dipole}
\end{align}
where $p_3$ is a Poincar\'e generator in $\alg{so}(2,4)$ and 
$h_4$ is a Cartan generator in $\alg{so}(6)$\,.  
In fact, the classical $r$-matrix leads to 
the deformed metric and NS-NS two form given by, respectively,  
\begin{align}
ds^2&=\frac{-(dx^0)^2+(dx^1)^2+(dx^2)^2+G^{-1}_1(dx^3)^2+dz^2}{z^2} \nln
&\quad 
+d\mu_1^2+G^{-1}_1 \mu_1^2\,d\psi_1^2 
+\sum_{i=2,3}(d\mu_i^2+\mu_i^2\, d\psi_i^2) \,, \nln 
B_2&= \al_1\,G^{-1}_1 z^{-2} \mu_1^2\,dx^3\wedge  d\psi_1 \,. 
\label{dipole-1para}
\end{align}
Here the scalar function $G_1$ is defined as  
\begin{align}
G_1\equiv 1+\al_1^2\,\mu_1^2\,z^{-2} \,. 
\end{align}
We have also used the coordinates of S$^5$ given in \eqref{mu123}\,. 
Indeed, the background with \eqref{dipole-1para} perfectly agrees with 
(6.30) in \cite{Imeroni}. Thus it has been shown that the classical $r$-matrix 
(\ref{1-para-dipole}) corresponds to this TsT transformation. 
This result gives a further support for the gravity/CYBE correspondence.

\subsection{A three-parameter case}
\label{subsec:4.2} 

The next is to consider a three-parameter generation of 
the classical $r$-matrix (\ref{dipole-1para})\,. 

\medskip 

Since $\alg{so}(6)$ has the three Cartan generators $h_4,h_5$ and $h_6$\,, 
one may consider the following three-parameter generalization;  
\begin{align}
r=\frac{i}{4\eta}\, p_3\wedge (\al_1\, h_4+\al_2\, h_5+\al_3\, h_6)\,, 
\label{3-dipole}
\end{align}
with three real constants $\al_1,\al_2$ and $\al_3$\,. 
Note that, when $\al_2=\al_3=0$\,, it reduces to the one-parameter case (\ref{dipole-1para})\,. 
After some computations, the resulting metric and NS-NS two-form are given by 
\begin{align}
ds^2&=\frac{-(dx^0)^2+(dx^1)^2+(dx^2)^2+G^{-1}_3(dx^3)^2+dz^2}{z^2} \nln
&\quad  +\sum_{i=1}^3(d\mu_i^2+\mu_i^2\, d\psi_i^2) 
-G^{-1}_3z^{-2} \Bigl(\sum_{i=1}^3\al_i\,\mu_i^2\, d\psi_i\Bigr)^2\,, \nln 
B_2&= z^{-2}G^{-1}_3\,dx^3\wedge\Bigl(\sum_{i=1}^3\al_i\,\mu_i^2\, d\psi_i\Bigr) \,.  
\label{dipole-3para}
\end{align}
Here the scalar function $G_3$ is defined as  
\begin{align}
G_3 \equiv 1+z^{-2} (\al_1^2\,\mu_1^2+\al_2^2\,\mu_2^2+\al_3^2\,\mu_3^2)\,. 
\end{align}
When $\alpha_1=\alpha_2=\alpha_3=0$\,, $G_3$ becomes $1$ and $B_2$ vanishes. 

\medskip 

The deformed background with \eqref{dipole-3para} can also be obtained by 
a sequence of three TsT-transformations:   
1) $(x^3, \psi_1)_{\al_1}$\,, 2) $(x^3, \psi_2)_{\al_2}$ and 3) $(x^3, \psi_3)_{\al_3}$\,. 
The derivation is straightforward but the result has not been listed in \cite{Imeroni}. 
Hence we have derived the explicit expressions in Appendix \ref{app:C}. 
The resulting metric and NS-NS two-form completely agree with 
the expressions in (\ref{dipole-3para}).

\section{Conclusion and Discussion}
\label{sec:5}

We have presented further examples of the gravity/CYBE correspondence. 
In the previous works, classical $r$-matrices have been composed of generators of 
only one of either $\mathfrak{so}(2,4)$ or $\mathfrak{so}(6)$\,. 
In this paper, we consider some examples of $r$-matrices with both of them. 
The $r$-matrices of this kind contain (generalized) Schr\"odinger spacetimes and gravity duals of 
dipole field theories. It is known that the generalized Schr\"odinger spacetimes can also be 
obtained via a certain class of TsT transformations called null Melvin twists. 
The metric and NS-NS two-form are reproduced by following the Yang-Baxter sigma-model description. 
Thus, this agreement shows that these backgrounds are classically integrable at the bosonic 
sigma-model level in the sense that the Lax pairs exist. 
We have to make efforts to prove this statement including the fermionic sector. 

\medskip

So far, the gravity/CYBE correspondence has been confirmed only for the metric (in the string frame) 
and NS-NS two-from. Hence the remaining task is to check the dilaton and R-R sector. 
In general, the dilaton is not constant and the R-R sector is also very complicated. 
Thus it does not seem so easy to check the remaining sector by explicitly evaluating 
the operator insertion into the classical action. However, the Schr\"odinger spacetime 
with a one-parameter is a bit special. Although the metric is deformed and the NS-NS 
two-form is newly turned on, the dilaton is still constant and the R-R sector is not 
modified. This result indicates that the Schr\"odinger spacetime would be a nice laboratory 
to check the gravity/CYBE correspondence at the full-sector level. 
We hope that we could report the result in the near future \cite{future}.  

\medskip 

There are various applications of the results presented here. 
An exciting issue is to consider applications to integrable deformations of type IIA string theory on AdS$_4 \times \mathbb{C}$P$^3$ \cite{IIA}. 
This system is dual to the $\mathcal{N}$=6 $SU(N) \times SU(N)$ 
Chern-Simons matter system in three dimensions \cite{ABJM}. This system was proposed by 
Aharony-Bergman-Jafferis-Maldacena (ABJM) \cite{ABJM} and it is often called the ABJM model. 
Hence the deformations of the string-theory side should correspond to deformations 
of the ABJM model, and there should be the associated classical $r$-matrices in the spirit of 
the gravity/CYBE correspondence. 

\medskip 

In particular, the most significant one is a non-relativistic limit of the ABJM model \cite{NR-ABJM}. 
This system preserves a super Schr\"odinger symmetry and the internal symmetry is also revealed. 
However, the gravity dual for this non-relativistic ABJM model has not been constructed 
yet\footnote{Early trials to look for the gravity dual \cite{OP,Eoin} have supposed 
the five-dimensional Schr\"odinger geometries. This is a possible line of approach, 
but as another possibility one of the internal directions may be external by acting a classical $r$-matrix. 
This is what we have in mind.}. 
Thus, it may be interesting to try to find out the gravity dual by employing the Yang-Baxter deformations 
of type IIA string theory. 
The information on the isometry obtained in \cite{NR-ABJM} 
would be a key ingredient to find out the corresponding classical $r$-matrix. 
We hope that our results shed light on the gravity dual for the non-relativistic ABJM model.

\subsection*{Acknowledgments}

We are very grateful to Io Kawaguchi for collaboration at the early stage of this work. 
This work is supported in part by the JSPS Japan-Hungary 
Research Cooperative Program.

\appendix

\section*{Appendix}

\section{Notation and convention \label{app:notation} }

We summarize here the notation and convention of 
the $\mathfrak{so}(2,4)$ and $\mathfrak{so}(6)$ generators, 
and a coset representation of AdS$_5\times$S$^5$\,. 

\subsubsection*{The gamma matrices}

In the following, we use the gamma-matrices represented by   
\begin{align}
& \ga_1=\begin{psmallmatrix}
0&0&0&-1\\
0&0&1&0\\
0&1&0&0\\
-1&0&0&0
\end{psmallmatrix} \,, 
\quad 
\ga_2=
\begin{psmallmatrix}
0&0&0&i\\
0&0&i&0\\
0&-i&0&0\\
-i&0&0&0 
\end{psmallmatrix}\,, 
\quad 
\ga_3=
\begin{psmallmatrix}
0&0&1&0\\
0&0&0&1\\
1&0&0&0\\
0&1&0&0
\end{psmallmatrix}\,, 
\nln  
& \ga_0=i\ga_4=
\begin{psmallmatrix}
0&0&1&0\\
0&0&0&-1\\
-1&0&0&0\\
0&1&0&0
\end{psmallmatrix}\,, 
\quad 
\ga_5= i\ga_1\ga_2\ga_3\ga_0=   
\begin{psmallmatrix}
1&0&0&0\\
0&1&0&0\\
0&0&-1&0\\
0&0&0&-1
\end{psmallmatrix}\,. 
\end{align}
To describe the $\alg{so}(2,4)$ and $\alg{so}(6)$ subalgebras of the $\alg{psu}(2,2|4)$ 
superalgebra, it is necessary to introduce the following $8\times 8$ gamma matrices: 
\begin{align}
\ga_\mu^{a}&=\begin{pmatrix} \ga_\mu & 0 \\ 0&0 \end{pmatrix}\,, & 
\ga_5^{a}&=\begin{pmatrix} \ga_5 & 0 \\ 0&0 \end{pmatrix} &
&\text{with} \qquad \mu=1,2,3,0\,, \nln
\ga_i^{s}&=\begin{pmatrix} 0&0 \\ 0& \ga_i \end{pmatrix}\,, &  
\ga_5^{s}&=\begin{pmatrix} 0&0 \\ 0& \ga_5 \end{pmatrix} &
&\text{with} \qquad i=1,2,3,4\,. \nonumber 
\end{align}
Here each block of the matrices is a $4\times4$ matrix. 

\subsubsection*{The bosonic generators}

Then, the Lie algebras $\alg{so}(2,4)$ and $\alg{so}(6)$ are spanned by the bases:  
\begin{align}
\alg{so}(2,4)&= {\rm span}_{\mathbb{R}}\{~\ga^a_\mu\,,\ga^a_5\,,
m_{\mu\nu}=\tfrac{1}{4}[\ga^a_\mu,\ga^a_\nu]\,,
m_{\mu5}=\tfrac{1}{4}[\ga^a_\mu,\ga^a_5]~|~~\mu,\nu=1,2,3,0~\}\,, \nln 
\alg{so}(6)&= {\rm span}_{\mathbb{R}}\{~\ga^s_i\,,\ga^s_5\,,
n_{ij}=\tfrac{1}{4}[\ga^s_i,\ga^s_j]\,,
n_{i5}=\tfrac{1}{4}[\ga^s_i,\ga^s_5]~|~~i,j=1,2,3,4~\}\,. 
\end{align}
Note that the subalgebras $\alg{so}(1,4)$ and $\alg{so}(5)$ 
are generated by 
\begin{align}
\alg{so}(1,4)&= {\rm span}_{\mathbb{R}}\{~
m_{\mu\nu}\,, m_{\mu5}~|~~\mu,\nu=1,2,3,0~\}\,, \nln 
\alg{so}(5)&= {\rm span}_{\mathbb{R}}\{~
n_{ij}\,,n_{i5}~|~~i,j=1,2,3,4~\}\,. 
\end{align}

\medskip 

For the coset construction of AdS$_5$ with the Poincar\'e coordinates, 
the following basis of $\alg{so}(2,4)$ is convenient;  
\begin{align}
\alg{so}(2,4)={\rm span}_{\mathbb{R}}\{~p_\mu\,,k_\mu\,,h_1\,,h_2\,,h_3\,, 
m_{13}\,,m_{10}\,,m_{23}\,,m_{20}~|~\mu=0,1,2,3~\}\,, 
\end{align}
where the Cartan generators $h_1, h_2, h_3$ and $p_\mu\,,k_\mu$ are given by 
\begin{align}
h_1&=2i m_{12}={\rm diag}(-1,1,-1,1,0,0,0,0)\,, & 
p_\mu&=\tfrac{1}{2}\ga^a_\mu-m_{\mu5}\,, \nln 
h_2&=2 m_{30}={\rm diag}(-1,1,1,-1,0,0,0,0)\,, &
k_\mu&=\tfrac{1}{2}\ga^a_\mu+m_{\mu5}\,, \nln 
h_3&=\ga_5^a={\rm diag}(1,1,-1,-1,0,0,0,0)\,.  
\end{align}
Note that the generators $p_\mu$ and $k_\mu$ commute each other, 
\begin{align}
[p_\mu,p_\nu]=[k_\mu,k_\nu]=[p_\mu,k_\nu]=0
\qquad \text{for} \qquad \mu, \nu=0,1,2,3\,. 
\end{align}

\medskip 

On the other hand, the Cartan generators of $\alg{so}(6)$ read
\begin{align}
h_4&=2i n_{12}={\rm diag}(0,0,0,0,-1,1,-1,1)\,, \nln 
h_5&=2i n_{34}={\rm diag}(0,0,0,0,-1,1,1,-1)\,,  \nln 
h_6&=\ga_5^s={\rm diag}(0,0,0,0,1,1,-1,-1)\,. 
\end{align}

\subsubsection*{A parameterization of the bosonic group elements}

We are now ready to parametrize bosonic group elements of $PSU(2,2|4)$\,. 
The group elements of  $SO(2,4)$ and $SO(6)$ are parametrized as  
\begin{align}
g_a&= \exp\bigl(x^1p_1+x^2p_2+x^3p_3+x^0p_0 \bigr)
\exp \Bigl(\frac{1}{2} \log z \ga^a_5\Bigr)
\nln
&= \exp\bigl(x^1p_1+x^2p_2+x^+p_++x^-p_- \bigr)
\exp \Bigl(\frac{1}{2} \log z \ga^a_5\Bigr) \quad \in SO(2,4)\,,  
\nln
g_s &=\exp(\psi^1h_4+\psi^2h_5+\psi^3h_6)\exp(-\ze n_{13})
\exp\Bigl(-\frac{i}{2}r\ga^s_1\Bigr)\quad \in SO(6)\,. \nonumber 
\end{align}
Here the light-cone coordinates and the associated 
generators are given by 
\begin{align}
x^\pm=\frac{x^0\pm x^3}{\sqrt{2}}\,, \qquad 
p_\pm =\frac{p_0 \pm p_3}{\sqrt{2}}\,. 
\label{app:lc}
\end{align}
Thus, a bosonic element $g$ of $PSU(2,2|4)$ is represented by 
\begin{align}
g=g_a g_s\quad \in SO(2,4)\times SO(6)\subset PSU(2,2|4)\,. 
\end{align}

\subsubsection*{Coset projector}
To derive the metric of AdS$_5\times$S$^5$ from the left-invariant one-form, 
\begin{align}
A=g^{-1}d g\quad \in \alg{so}(2,4)\oplus\alg{so}(6)\,,  
\label{app:current}
\end{align}
it is necessary to introduce the coset projector;   
\begin{align}
P_2:~~~ \alg{so}(2,4)\oplus\alg{so}(6) ~~\longrightarrow~~
\frac{\alg{so}(2,4)}{\alg{so}(1,4)} \oplus \frac{\alg{so}(6)}{\alg{so}(5)}\,. 
\end{align}
For any $x\in \alg{so}(2,4)\oplus\alg{so}(6)$\,, it is explicitly defined as 
\begin{align}
P_2(x)=\frac{1}{4}\bigl(\eta^{\mu\nu} \ga_\mu^a\Tr[\ga_\nu^a x]
+\ga_5^a\Tr[\ga_5^a x] +\de^{ij} \ga_i^s\Tr[\ga_j^s x] \bigr)\,, 
\label{app:proj}
\end{align}
where the range of the indices are $\mu,\nu=0,1,2,3$ and $i,j=1,2,3,4,5$\,. 
Then the four-dimensional Minkowski metric is given by 
\begin{align}
\eta^{\mu\nu}={\rm diag}(-1,1,1,1)\,. 
\end{align}

\subsubsection*{The AdS$_5\times$S$^5$ metric}
 
With the left-invariant one-form \eqref{app:current} and
the coset projector \eqref{app:proj}, the AdS$_5\times$S$^5$ metric 
can be reproduced as 
\begin{align}
\str[AP_2(A)]=ds^2_{\rm AdS_5}+ds^2_{\rm S^5}\,, 
\end{align}
where $ds^2_{\rm AdS_5}$ and $ds^2_{\rm S^5}$ are the metrics of 
AdS$_5$ and S$^5$ respectively,  
\begin{align}
ds^2_{\rm AdS_5}&= \frac{1}{z^2}\bigl(-2dx^+dx^-+(dx^1)^2+(dx^2)^2+dz^2\bigr)\,, \nln 
ds^2_{\rm S^5}&=dr^2+\sin^2r \bigl(d\ze^2+ \cos^2\ze (d\psi_1)^2
+ \sin^2\ze (d\psi_2)^2\bigr)+\cos^2r (d\psi_3)^2\,. 
\label{app:S5metric}
\end{align}
This is the standard representation of the AdS$_5\times$S$^5$ metric. 
It is noted that the ranges of S$^5$ coordinates are restricted as follows:   
\begin{align}
0 \leq r \leq \frac{\pi}{2}\,, \qquad 
0 \leq \zeta \leq \frac{\pi}{2}\,, \qquad 
0 \leq \psi_i \leq 2\pi \quad (i=1,2,3)\,. 
\end{align}

\subsubsection*{The coordinate systems of S$^5$}

When we argue the Schr\"odinger geometry, 
it is more convenient to rewrite the S$^5$ metric \eqref{app:S5metric}  as 
an S$^1$-fibration over $\mathbb{C}$P$^2$\,: 
\begin{align} 
ds^2_{\rm S^5}&=(d\chi+\omega)^2 +ds^2_{\rm \mathbb{C}P^2}\,, \nln 
ds^2_{\rm \mathbb{C}P^2}&= d\mu^2+\sin^2\mu
\bigl(\Sigma_1^2+\Sigma_2^2+\cos^2\mu\Sigma_3^2\bigr)\,,  
\label{app:S1overCP2}
\end{align}
where $\chi$ is the S$^1$-fiber coordinate and $\omega$ is the one-form
potential of the K\"ahler form on $\mathbb{C}$P$^2$\,. 
The symbols $\Sigma_i ~(i=1,2,3)$ and $\omega$ are defined by 
\begin{align}
\Sigma_1&= \tfrac{1}{2}(\cos\psi d\theta +\sin\psi\sin\theta d\phi)\,, \nln 
\Sigma_2&= \tfrac{1}{2}(\sin\psi d\theta -\cos\psi\sin\theta d\phi)\,, \nln 
\Sigma_3&= \tfrac{1}{2}(d\psi +\cos\theta d\phi)\,, 
\qquad 
\omega=\sin^2\mu \Sigma_3\,. 
\end{align}
The coordinate transformation from \eqref{app:S5metric} to  
\eqref{app:S1overCP2} is given by 
\begin{align} 
\psi_1&= \chi+\tfrac{1}{2}(\psi+\phi)\,, \qquad  r= \mu\,, \nln 
\psi_2&= \chi+\tfrac{1}{2}(\psi-\phi)\,, \qquad \ze= \tfrac{1}{2}\theta\,, \nln  
\psi_3&=\chi\,.   
\end{align}

\section{The T-duality rules} 
\label{app:Tdual} 

The rules of T-duality \cite{Buscher,T-duality} are summarized here.  
We basically follow the rules of \cite{T-duality}. 

\medskip 

The transformation rules between type IIB and type IIA supergravities 
are listed below. 
Note that the T-duality is performed for the $y$-direction and the other coordinates are denoted by $a,b,a_i~(i=1,\ldots)$\,. 
The fields of type IIB supergravity are the metric $g_{\mu\nu}$\,, NS-NS two-form $B_2$\,, dilaton $\Phi$\,, 
R-R gauge fields $C^{(2n)}$\,. The ones of type IIA supergravity are denoted with the tilde, 
the metric $\tilde{g}_{\mu\nu}$\,, NS-NS two-form $\tilde{B}_2$\,, dilaton $\tilde{\Phi}$\,, 
and R-R gauge fields $\tilde{C}^{(2n+1)}$\,. 

\paragraph{From type IIB to type IIA}  
\begin{align}
\tilde g_{yy}&= \frac{1}{g_{yy}}\,, \qquad \tilde g_{ay}=\frac{B_{ay}}{g_{yy}}\,, 
\qquad \tilde g_{ab}=g_{ab}-\frac{g_{ya}g_{yb}-B_{ya}B_{yb}}{g_{yy}}\,, \el \\
\tilde B_{ay}&=\frac{g_{ay}}{g_{yy}}\,, \qquad \tilde B_{ab}=B_{ab}-\frac{g_{ya}B_{yb}-B_{ya}g_{yb}}{g_{yy}}\,,
\qquad 
\tilde \Phi=\Phi-\frac{1}{2}\ln g_{yy}\,, \nonumber \\ 
\tilde{C}^{(2n+1)}_{a_1\cdots a_{2n+1}} 
&= - C^{(2n+2)}_{a_1\cdots a_{2n+1}y} - (2n+1) B^{~}_{y[a_1}C^{(2n)}_{a_2\cdots a_{2n+1}]} 
+2n(2n+1)\frac{B^{~}_{y[a_1}g^{~}_{a_2 |y|} C^{(2n)}_{a_3\cdots a_{2n+1}]y}}{g_{yy}}\,, \nonumber \\ 
\tilde{C}^{(2n+1)}_{a_1\cdots a_{2n}y} &= C^{(2n)}_{a_1\cdots a_{2n}} 
+ 2n \frac{g^{~}_{y[a_1}C^{(2n)}_{a_2\cdots a_{2n}]y}}{g_{yy}}\,. 
\end{align}
where the anti-symmetrization for indices is defined as, for example,  
\[
A_{[a}B_{b]} \equiv \frac{1}{2}\left(A_a B_b - A_b B_a \right)\,. 
\]
The symbol $|y|$ inside the anti-symmetrization means that the indices other than the index $y$ are anti-symmetrized. 

\paragraph{From type IIA to type IIB}

\begin{align}
g_{yy}&= \frac{1}{\tilde{g}_{yy}}\,, \qquad g_{ay}=\frac{\tilde{B}_{ay}}{\tilde{g}_{yy}}\,, 
\qquad g_{ab} = \tilde{g}_{ab}-\frac{\tilde{g}_{ya}\tilde{g}_{yb}-\tilde{B}_{ya}\tilde{B}_{yb}}{\tilde{g}_{yy}}\,, \el \\
B_{ay}&=\frac{\tilde{g}_{ay}}{\tilde{g}_{yy}}\,, \qquad B_{ab} = \tilde{B}_{ab} - \frac{\tilde{g}_{ya}\tilde{B}_{yb} 
- \tilde{B}_{ya}\tilde{g}_{yb}}{\tilde{g}_{yy}}\,, \qquad 
\Phi = \tilde{\Phi}-\frac{1}{2}\ln \tilde{g}_{yy}\,, \nonumber \\ 
C^{(2n)}_{a_1\cdots a_{2n}} &= \tilde{C}^{(2n+1)}_{a_1\cdots a_{2n}y} - 2n \tilde{B}^{~}_{y[a_1}\tilde{C}^{(2n-1)}_{a_2\cdots a_{2n}]} 
+2n(2n-1)\frac{\tilde{B}^{~}_{y[a_1}\tilde{g}^{~}_{a_2|y|}\tilde{C}^{(2n-1)}_{a_3\cdots a_{2n}]y}}{\tilde{g}_{yy}}\,, \nonumber \\ 
C^{(2n)}_{a_1\cdots a_{2n-1}y} &= - \tilde{C}^{(2n-1)}_{a_1\cdots a_{2n-1}} 
- (2n-1)\frac{\tilde{g}^{~}_{y[a_1}\tilde{C}^{(2n-1)}_{a_2\cdots a_{2n-1}]y}}{\tilde{g}_{yy}}\,. 
\end{align}

\section{Derivation of dipole deformations of AdS$_5\times$S$^5$}
\label{app:C}

Let us here derive gravity duals for dipole theories by performing 
TsT transformations for AdS$_5\times$S$^5$\,. 
In the following, we will concentrate on the NS-NS sector. 
The T-duality rules we utilize is summarized in Appendix \ref{app:Tdual}.

\subsection{One-parameter deformation}

First of all, we consider a one-parameter deformation of AdS$_5\times$S$^5$\,. 
The starting point is the following metric, NS-NS two-form $B_2$ and dilaton $\Phi$~:
\begin{eqnarray}
ds^2 &=& \frac{-(dx^0)^2+(dx^1)^2+(dx^2)^2 + (dx^3)^2+dz^2}{z^2} 
+ \sum_{i=1}^3(d\mu_i^2 + \mu_i^2\,d\psi_i^2)\,, \nonumber \\ 
&& B_2 =0\,, \qquad \Phi = \Phi_0~~(\mbox{const.})\,.
\end{eqnarray}

\medskip 

For this background, we will perform a TsT transformation ($x^3$,$\psi_1$)$_{\alpha_1}$\,. 
For this purpose, the relevant part is 
\begin{eqnarray}
ds^2 = \frac{1}{z^2} (dx^3)^2 + \mu_1^2\, d \psi_1^2\,, \quad B_2=0\,, \quad \Phi=\Phi_0\,.
\end{eqnarray}
The first step is a T-duality for the $x^3$-direction. 
The resulting background is given by 
\begin{eqnarray}
d\tilde{s} {}^2 &=& z^2 (d\tilde{x} {}^3)^2 + \mu_1^2\,d\psi_1^2\,, \quad B_2=0\,, 
\quad \Phi=\Phi_0-\frac{1}{2}\ln\left(\frac{1}{z^2}\right)\,.
\label{TsT1-1}
\end{eqnarray}
Then, by shifting $\psi_1$ as $\psi_1 = \tilde{\psi}_1-\alpha_1\, \tilde{x}^3$\,, 
the background (\ref{TsT1-1}) is rewritten as 
\begin{eqnarray}
d\tilde{s}^2 &=& (z^2+\mu_1^2\,\alpha_1^2)(d\tilde{x}^3)^2 
+ \mu_1^2\, d\tilde{\psi}_1^2 -2\mu_1^2\,\alpha_1\,d\tilde{\psi}_1d\tilde{x}^3\,, \nonumber \\ 
&& \tilde{B}_2 =0\,, \qquad \tilde{\Phi} = \Phi_0 + \frac{1}{2}\ln z^2\,.
\end{eqnarray}
Finally, a T-duality is performed for the $\tilde{x}^3$-direction. 
Together with the undeformed part, the resulting background is given by 
\begin{eqnarray}
ds^2 &=& \frac{-(dx^0)^2+(dx^1)^2+(dx^2)^2 + G_1^{-1} (dx^3)^2+dz^2}{z^2} 
+ \sum_{i=1}^3(d\mu_i^2 + \mu_i^2\,d\psi_i^2)\,, \nonumber \\ 
&& B_2 = \alpha_1 G_1^{-1}z^{-2}\mu_1^2\,dx^3\wedge d\psi_1\,, \qquad {\rm e}^{2\Phi} = {\rm e}^{2\Phi_0}\,G_1^{-1}\,, 
\label{TsT1-3}
\end{eqnarray}
Here we have removed the tilde from $\psi_1$ and a scalar function $G_1(z)$ is defined as 
\[
G_1 (z) \equiv 1 + \alpha_1^2\,\mu_1^2\,z^{-2}\,.
\]

\subsection{Two-parameter deformation}

We will next consider the second TsT transformation $(x^3,\psi_2)_{\alpha_2}$ for the background (\ref{TsT1-3})\,. 
The relevant part is 
\begin{eqnarray}
ds^2 &=& G_1^{-1}\frac{(dx^3)^2}{z^2} + G_1^{-1}\mu_1^2\,d\psi_1^2 + \mu_2^2\,d\psi_2^2\,, \nonumber \\ 
&& B_{\psi_1 x^3} = -\frac{\mu_1^2\,\alpha_1}{z^2+\mu_1^2\alpha_1}\,, 
\qquad \Phi = \Phi_0 -\frac{1}{2}\ln G_1\,.
\end{eqnarray}
We first perform a T-duality for the $x^3$-direction. 
The above part is rewritten as 
\begin{eqnarray}
d\tilde{s}^2 &=& z^2 G_1\,(d\tilde{x}^3)^2 + \mu_1^2\,d\psi_1^2 + \mu_2^2\,d\psi_2^2 
-2\mu_1^2\,\alpha_1\,d\psi_1d\tilde{x}^3\,, \nonumber \\ 
&& \tilde{B}_2 = 0\,, \qquad \tilde{\Phi} = \Phi_0 + \frac{1}{2}\ln z^2\,. 
\label{TsT2-1}
\end{eqnarray}
Then, by shifting $\psi_2$ as $\psi_2 = \tilde{\psi}_2 -\alpha_2\,\tilde{x}^3$\,, 
the background (\ref{TsT2-1}) is rewritten as 
\begin{eqnarray}
d\tilde{s}^2 &=& (z^2 G_1 +\alpha_2^2\,\mu_2^2)\,(d\tilde{x}^3)^2 + \mu_1^2\,d\psi_1^2 + \mu_2^2\,d\tilde{\psi}_2^2 
-2\mu_1^2\,\alpha_1\,d\psi_1d\tilde{x}^3 -2\mu_2^2\,\alpha_2\,d\tilde{\psi}_2d\tilde{x}^3\,, \nonumber \\ 
&& \tilde{B}_2 = 0\,, \qquad \tilde{\Phi} = \Phi_0 + \frac{1}{2}\ln z^2\,. 
\label{TsT2-2}
\end{eqnarray}
Finally, we perform a T-duality for the $\tilde{x}^3$-direction. 
Together with the undeformed part,the resulting background is given by 
\begin{eqnarray}
ds^2 &=& \frac{-(dx^0)^2+(dx^1)^2+(dx^2)^2 + G_2^{-1} (dx^3)^2+dz^2}{z^2} 
\,, \nonumber \\ 
&& \qquad  + \sum_{i=1}^3(d\mu_i^2 + \mu_i^2\,d\psi_i^2)-z^{-2}G_2^{-1}(\alpha_1\,\mu_1^2\,d\psi_1 
+ \alpha_2\,\mu_2^2\,d\psi_2)^2\,, \nonumber \\
&& B_2 = z^{-2}G_2^{-1}dx^3\wedge (\alpha_1\,\mu_1^2\,d\psi_1+\alpha_2\,\mu_2^2\,d\psi_2)\,, 
\qquad {\rm e}^{2\Phi} = {\rm e}^{2\Phi_0}\,G_2^{-1}\,, 
\label{TsT2-3}
\end{eqnarray}
Here we have removed the tilde from $\psi_2$ and a scalar function $G_2(z)$ is defined as 
\[
G_2 (z) \equiv 1 + z^{-2}(\alpha_1^2\,\mu_1^2 + \alpha_2^2\,\mu_2^2)\,.
\]

\subsection{Three-parameter deformation}

Finally, let us perform the third TsT transformation $(x^3,\psi_3)_{\alpha_3}$ for the background (\ref{TsT2-3})\,. 
The relevant part is given by 
\begin{eqnarray}
ds^2 &=& G_2^{-1}\frac{(dx^3)^2}{z^2} + \sum_{i=1}^3 \mu_i^2\,d\psi_i^2 -z^{-2}G_2^{-1}\,(\alpha_1\,\mu_1^2\,d\psi_1 
+ \alpha_2\,\mu_2^2\,d\psi_2)^2\,, \nonumber \\ 
&& B_{\psi_1 x^3} = -\frac{\mu_1^2\,\alpha_1}{z^2}\,G_2^{-1}\,, \quad 
 B_{\psi_2 x^3} = -\frac{\mu_2^2\,\alpha_2}{z^2}\,G_2^{-1}\,, \qquad \Phi = \Phi_0 -\frac{1}{2}\ln G_2\,. 
\end{eqnarray}
By performing a T-duality for the $x^3$-direction, the above part is recast into 
\begin{eqnarray}
d\tilde{s}^2 &=& z^2\,G_2\,(d\tilde{x}^3)^2 + \sum_{i=1}^3\mu_i^2\,d\psi_i^2 
-2\alpha_1\,\mu_1^2\,d\psi_1 d\tilde{x}^3 -2\alpha_2\,\mu_2^2\,d\psi_2 d\tilde{x}^3\,, \nonumber \\ 
&& \tilde{B}_2 =0\,, \qquad \tilde{\Phi} = \Phi_0 + \frac{1}{2}\ln z^2\,. 
\label{TsT3-1}
\end{eqnarray}
Then, by shifting $\psi_3$ as $\psi_3 = \tilde{\psi}_3 -\alpha_3\,\tilde{x}^3$\,, 
the background (\ref{TsT3-1}) is rewritten as 
\begin{eqnarray}
d\tilde{s}^2 &=& (z^2 G_2 +\alpha_3^2\,\mu_3^2)\,(d\tilde{x}^3)^2 + \mu_1^2\,d\psi_1^2 + \mu_2^2\,d\psi_2^2 
+ \mu_3^2\,d\tilde{\psi}_3^2 \nonumber \\ 
&& -2\mu_1^2\,\alpha_1\,d\psi_1d\tilde{x}^3  -2\mu_2^2\,\alpha_2\,d\psi_2 d\tilde{x}^3 
-2\mu_3^2\,\alpha_3\,d\tilde{\psi}_3 d\tilde{x}^3\,, \nonumber \\ 
&& \tilde{B}_2 = 0\,, \qquad \tilde{\Phi} = \Phi_0 + \frac{1}{2}\ln z^2\,. 
\label{TsT3-2}
\end{eqnarray}
Finally, we perform a T-duality for the $\tilde{x}^3$-direction. 
Together with the undeformed part,the resulting background is given by 
\begin{eqnarray}
ds^2 &=& \frac{-(dx^0)^2+(dx^1)^2+(dx^2)^2 + G_3^{-1} (dx^3)^2+dz^2}{z^2} 
\,, \nonumber \\ 
&& \qquad  + \sum_{i=1}^3(d\mu_i^2 + \mu_i^2\,d\psi_i^2)-z^{-2}G_3^{-1}
\left(\sum_{i=1}^3 \alpha_i\,\mu_i^2\,d\psi_i\right)^2\,, \nonumber \\
&& B_2 = z^{-2}G_3^{-1}dx^3\wedge \left(\sum_{i=1}^3 \alpha_i\,\mu_i^2\,d\psi_i\right)\,, 
\qquad {\rm e}^{2\Phi} = {\rm e}^{2\Phi_0}\,G_3^{-1}\,, 
\label{TsT3-3}
\end{eqnarray}
Here we have removed the tilde from $\psi_3$ and a scalar function $G_3(z)$ is defined as 
\[
G_3 (z) \equiv 1 + z^{-2}\sum_{i=1}^3 \alpha_i^2\,\mu_i^2\,.
\]

\end{document}